# Large superconducting diode effect driven by edge states in twisted iron-chalcogenide Josephson junctions

Xiangyu Zeng[*,1,2], Renjie Zhang[*,1], Guoliang Guo[*,1], Zhuoqing Gao[2], Quanxin Hu[1], Haijiao Ji[1,3], Fazhi Yang[4], Xiaozhi Wang[2], Bo Gao[1], Noah F. Q. Yuan[1,3], Baiqing Lv[†,1,3,5], Xin Liu[‡,1], Hong Ding[§,1,6,7]

[1]*Tsung-Dao Lee Institute, Shanghai Jiao Tong University, Shanghai 200240, China*
[2]*College of Information Science and Electronic Engineering, Zhejiang University, Hangzhou, 310027, China*
[3]*School of Physics and Astronomy, Shanghai Jiao Tong University, Shanghai 200240, China*
[4]*Department of Physics, City University of Hong Kong, Kowloon, Hong Kong, 999077, China*
[5]*Zhangjiang Institute for Advanced Study, Shanghai Jiao Tong University, Shanghai 200240, China*
[6]*Hefei National Laboratory, Hefei 230088, China*
[7]*New Cornerstone Science Laboratory, Shanghai 201201, China*

* These authors contributed equally to this work
†baiqing@sjtu.edu.cn
‡phyliuxin@sjtu.edu.cn
§dingh@sjtu.edu.cn

**ABSTRACT**

The superconducting diode effect (SDE)—the unidirectional, dissipationless flow of supercurrent—is a critical element for future superconducting electronics[1–7]. Achieving high efficiency under zero magnetic field is a key requirement. The Josephson junction constitutes a versatile SDE platform for exploiting quantum materials that exhibit ferromagnetism[4], topology[8], or unconventional superconductivity[9]. However, a single two-dimensional material system that inherently offers these properties and allows for precise interface engineering, such as twisting, remains elusive. Here we report a record-high, field-free diode efficiency of ~30% in twist van der Waals Josephson heterostructures of the sign-change iron-chalcogenide superconductor $FeTe_{0.55}Se_{0.45}$ and the conventional transition-metal dichalcogenide superconductor $2H\text{-}NbSe_2$. The diode response shows a striking twist-angle dependence: the efficiency peaks at crystallographic alignment and collapses with a small misorientation of ~7°. Importantly, the twist-angle evolution of superconducting interference measurements reveals that efficient nonreciprocity arises from asymmetric edge supercurrents, whereas bulk transport suppresses the effect. These findings establish edge states as the driving mechanism of the unconventional SDE, linking it to exotic pairing and topology in multiband iron-based superconductors. Our findings reveal intricate physics involving novel pairing symmetry, magnetism, and topology in the multiband iron-based superconductor, and offer a new route to high-performance superconducting diodes.

**INTRODUCTION**

Iron-based superconductors (FeSCs) are a class of quantum materials distinguished by their multiband electronic structure and the rich interplay among symmetry breaking correlated phases, topology, and unconventional superconductivity[10]. The multiband nature gives rise to a variety of emergent phenomena, including sign-changing superconducting gaps between different Fermi surfaces[11–13], as well as nontrivial topological bands and boundary states[14–20]. Among the FeSC families, the iron-chalcogenides are particularly notable. They possess a two-dimensional (2D) Ch–Fe–Ch (Ch = chalcogen) layered structure[21], where superconductivity and ferromagnetism can coexist[22–24], and exhibit topological boundary states[14–16] alongside a unique $s\pm$ pairing symmetry[12,25]. These intertwined properties, combined with the highest superconducting transition temperatures in the FeSC family[26], highlight both their fundamental scientific interest and significant technological promise.

These unique properties suggest that iron-chalcogenides could enable a field-free superconducting diode effect (SDE) through spontaneous ferromagnetism[4], edge states[28–30] and frustrated Josephson coupling[31–33]. Meanwhile, recent progress in twistronics offers a powerful new dimension for engineering Josephson heterostructures: analogous to how moiré potentials reshape the electronic states of twisted graphene[34,35], twisted superconducting interfaces can strongly modulate Josephson coupling[9,36]. In FeSCs, the sign-changing gap across multiple bands makes such coupling particularly sensitive to the interface[31–33], offering a new route to superconducting diodes. Despite the versatility of FeSCs, the twist-engineered SDE have remained unexplored.

Here we demonstrate a highly efficient, field-free SDE in twisted heterostructures between the $s\pm$-wave superconductor $FeTe_{0.55}Se_{0.45}$ and the conventional $s$-wave superconductor 2H-$NbSe_2$[37]. We achieve diode efficiencies up to ~30%, the highest reported to date in Josephson junctions without an applied magnetic field. The effect shows a pronounced twist-angle dependence, peaking when the crystal axes are aligned and vanishing at small misorientations.

This is the first demonstration of twist engineering in iron-based topological superconductors. Twist-angle dependent superconducting interference measurements reveal that efficient nonreciprocity arises from asymmetric edge supercurrents, whereas bulk-dominated transport suppresses the effect. Our findings establish twist-controlled coupling between distinct superconductors as a new strategy for creating high-performance superconducting diodes. This platform also offers an unprecedented opportunity to investigate the unconventional roles of exotic quantum states on the SDE.

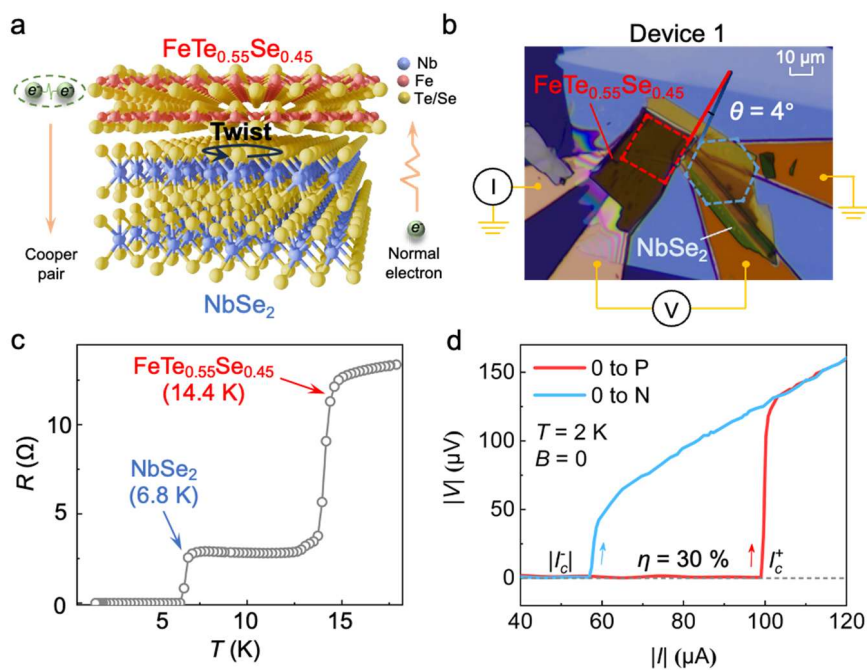

**Fig. 1 | Structure and zero-field superconducting diode effect (SDE) in a FeTe$_{0.55}$Se$_{0.45}$/NbSe$_2$ junction.** a, Schematic of the twisted FeTe$_{0.55}$Se$_{0.45}$/NbSe$_2$ junction illustrating the SDE. b, Optical micrograph of Device 1. The red square and blue hexagon denote the crystal lattices of FeTe$_{0.55}$Se$_{0.45}$ and NbSe$_2$, respectively, with their edges aligned along the crystallographic *a*-axes. The twist angle, defined as the relative orientation between the two *a*-axes, is ~4°. c, Temperature-dependent resistance of Device 1, showing two successive superconducting transitions that confirm the formation of a Josephson junction. d, Current–voltage characteristics of Device 1 at 2 K and zero magnetic field, displaying a clear SDE with an efficiency of ~30%. The curves were both obtained from current sweeps starting at zero bias.

**RESULTS AND DISCUSSIONS**

## 2.1 The Field-Free Superconducting Diode Effect

To investigate the transport properties and potential for a SDE in $FeTe_{0.55}Te_{0.45}$/$NbSe_2$ heterostructures, we fabricated a series of nine vertical vdW junctions (Device 1 to Device 9) by stacking exfoliated flakes of the $s\pm$-wave superconductor $FeTe_{0.55}Te_{0.45}$ and the conventional $s$-wave superconductor 2H-$NbSe_2$. All devices were prepared under near-identical conditions using flakes of comparable thickness. Optical images detailing the device geometry and the four-probe electrode configuration are shown in Fig. 1b and Supplementary Information S1. The full fabrication process is detailed in the Methods section.

The physical dimensions, particularly the flake thickness and the relative twist angle ($\theta$) between the layers, are crucial parameters governing the superconducting behavior in these 2D vdW heterostructures. Atomic force microscopy (AFM) measurements confirmed that all $FeTe_{0.55}Te_{0.45}$ and $NbSe_2$ flakes had thicknesses in the tens to hundreds of nanometers (Supplementary Information S2). Furthermore, transmission electron microscopy (TEM) confirmed the high quality of the interface, revealing a sharp junction free of oxidation (Supplementary Information S3). The twist angle $\theta$ represents a key degree of freedom in stacked 2D materials. We define $\theta = 0°$ as the alignment where the crystallographic $a$-axes of the four-fold symmetric $FeTe_{0.55}Te_{0.45}$ and the six-fold symmetric $NbSe_2$ are parallel (Fig. 1b). The precise twist angle for each device was determined using a combination of electron backscatter diffraction (EBSD) and natural cleavage edges (Supplementary Information S4).

Measurements of temperature-dependent resistance upon cooling revealed a characteristic two-step superconducting transition to zero resistance in each junction (Fig. 1c and Supplementary Information S5). The onset of the first transition at ~14 K corresponds to the bulk critical temperature ($T_c$) of $FeTe_{0.55}Te_{0.45}$, while the second transition at ~7 K matches the Tc of 2H-$NbSe_2$. This sequential behavior unambiguously confirms the successful formation of a vdW Josephson junction composed of two distinct superconducting materials.

With the Josephson coupling established, we next investigated the SDE by comparing the voltage-current ($V-I$) characteristics for both positive ($0 \rightarrow +I$) and negative ($0 \rightarrow -I$) current sweeps. We consistently observed a pronounced asymmetry between the positive ($I_c^+$) and negative ($I_c^-$) critical currents under zero magnetic field. This asymmetry confirms the presence of a strong field-free SDE, with a superconducting diode efficiency $\eta = |\frac{I_c^+ - |I_c^-|}{I_c^+ + |I_c^-|}| \times 100\%$ reaching up to 30.0% in Device 1.

From the symmetry perspective, the field-free SDE requires the simultaneous breaking of both inversion symmetry and time-reversal symmetry[38]. The vertical heterostructure geometry intrinsically breaks inversion symmetry, which is necessary, though not sufficient, for nonreciprocal transport. The requisite TRSB must be an intrinsic property of the materials or the interface. Previous studies have reported evidence of surface ferromagnetism coexisting with superconductivity in $FeTe_{0.55}Te_{0.45}$, even though the bulk material is not ferromagnetic. This ferromagnetism naturally provides the necessary TRSB, which is coupled with the broken inversion symmetry in the junction to enable the observed field-free SDE.

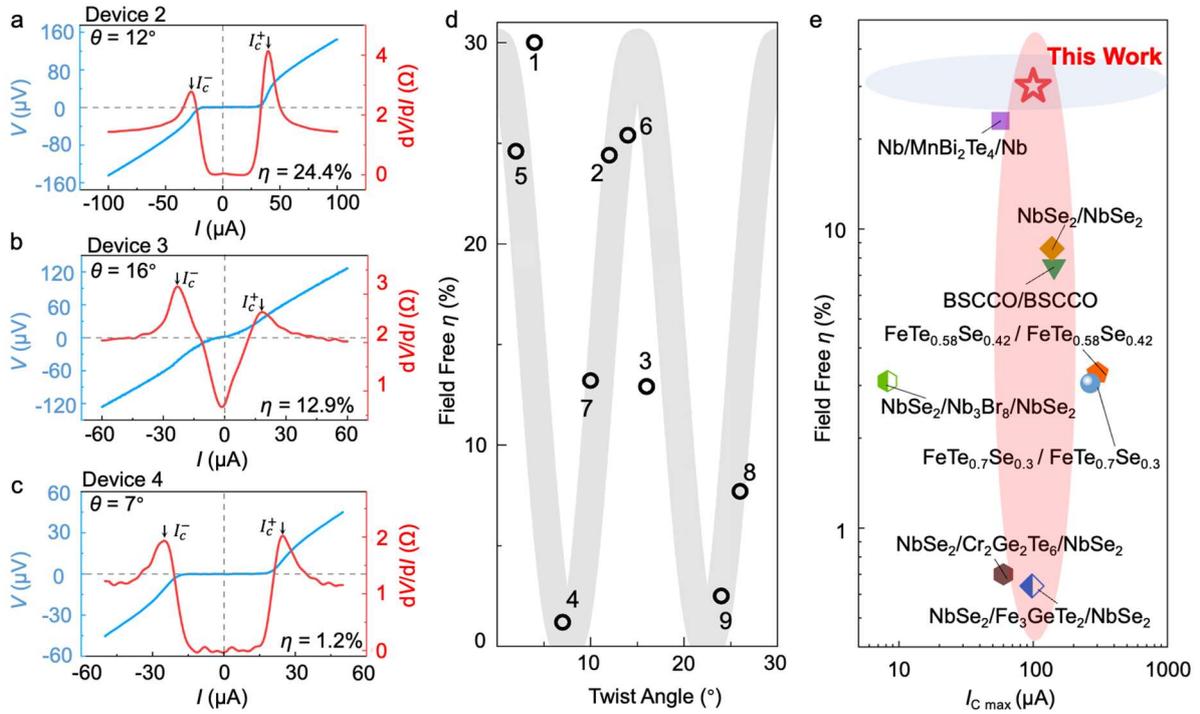

**Fig. 2 | Twist-angle dependence of the zero-field superconducting diode effect (SDE).** a–c, Current–voltage (blue) and differential-resistance (d$V$/d$I$) curves for Devices 2 (a), 3 (b), and 4 (c), measured at zero field. The current was swept from zero to positive bias for I > 0 and from zero to negative bias for I < 0. The twist angles of the devices are 12°, 16°, and 7°, with corresponding diode efficiencies of 24.4%, 12.9%, and 1.2%. d, Twist-angle dependence of the diode efficiency across Devices 1–9, measured at zero field. e, Comparison of the maximum zero-field diode efficiency ($\eta$) versus critical current ($I_{C\,max}$) for $FeTe_{0.55}Te_{0.45}/NbSe_2$ junctions

and other reported field-free SDE Josephson junctions (see Supplementary Information S6). The red dashed line marks $\eta$ = 25%. FeTe$_{0.55}$Te$_{0.45}$/NbSe$_2$ achieves a maximum efficiency of ~30% at $I_{C\,max}$ ~100 μA.

**2.2 Twist-Angle Control of the Field-free Superconducting Diode Effect**

Inspired by the concept of twistronics, we systematically explored the SDE across the range of twist angles ($\theta$) from 0° to 30° in the nine devices to elucidate the underlying mechanism supporting the large field-free SDE. Given the 6-fold rotational symmetry of NbSe$_2$ and the 4-fold symmetry of FeTe$_{0.55}$Te$_{0.45}$, the FeTe$_{0.55}$Te$_{0.45}$/NbSe$_2$ heterostructures exhibit a rotational period of 30°. All measured junctions exhibited a field-free SDE, as evidenced by the asymmetric $V-I$ curves and corresponding differential resistances (d$V$/d$I$) under zero magnetic field (Figs. 2a-2c and Supplementary Information S7). But more remarkably, we discovered a strong and non-trivial dependence of the diode efficiency on the twist angle $\theta$. For instance, the $V-I$ curves for Device 2 ($\theta$ = 3°) and Device 3 ($\theta$ = 15°) showed pronounced asymmetry, with efficiencies $\eta$ = 25.4% and $\eta$ = 12.9%, respectively (Figs. 2a-2b). In stark contrast, Device 4 ($\theta$ = 7°) exhibited an almost negligible efficiency of $\eta$ = 1.2% (Fig. 2c).

By plotting the efficiencies for the nine devices against their measured twist angles (Fig. 2d), a clear periodic dependence of $\eta$ emerges. The SDE efficiency peaks near $\theta$ = 0°, 15°, and 30°, and nearly vanishes at intermediate angles around 7° and 23°. Device 1 ($\theta$ = 4°) displays the strongest SDE, reaching a record-high efficiency of $\eta$ = 30%. This value, combined with a relatively large critical current (~100 μA, Fig. 2e), underscores the high performance and technological potential of the FeTe$_{0.55}$Te$_{0.45}$/NbSe$_2$ platform for high-power superconducting electronics. The observed angular modulation of the SDE suggests that the nonreciprocal transport is intimately linked to the twist angle, likely arising from the twist-controlled coupling between two distinct superconductors.

**2.3 Twist-Angle Dependence of Superconducting Interference Patterns**

After establishing the twist-angle-dependent field-free behavior, we next investigated the junctions' response to an applied in-plane magnetic field ($B$). Superconducting interference patterns (SIPs)—the modulation of the critical current with $B$ field—are instrumental in mapping the spatial distribution of the supercurrent (Supplementary Information S8), offering crucial insights into the unconventional Josephson coupling and the origin of the SDE.

We first examined devices exhibiting high diode efficiency, such as Device 2 ($\theta = 3°$). In striking contrast to the typical decaying Fraunhofer pattern characteristic of a uniform current density, Device 2 displays a robust, SQUID-like interference pattern with negligible $I_c$ amplitude decay over multiple oscillations (Fig. 3a and Supplementary Information S9 for Device 6). This non-decaying SIP strongly suggests the dominance of supercurrents localized at the edges of the junction.

The oscillation period ($B_p$) of the SIP allows us to estimate the effective magnetic area $S = \phi_0/B_p$, where $\phi_0$ is the magnetic flux quantum. From the measured period $B_p = 12$ Oe, we estimate $S \approx 1.6$ $\mu m^2$. This value closely matches the calculated geometric area $S_{phy} = L(d + \lambda_{FTS} + \lambda_{NS}) \approx 1.65$ $\mu m^2$, where $L \approx 50$ $\mu m$ is the junction length, $d < 1$ nm is the interlayer spacing, and $\lambda_{FTS} = 28$ nm and $\lambda_{NS} = 5$ nm are the London penetration depths of FeTe$_{0.55}$Te$_{0.45}$[24] and NbSe$_2$[39], respectively.

To definitively confirm the supercurrent distribution, we performed a Dynes-Fulton analysis[40] (Supplementary Information S10) on the SIP data. The resulting supercurrent density profile for Device 2 unequivocally shows that the asymmetric supercurrent is highly localized near the edges, with minimal contribution from the bulk (Fig. 3b). A similar edge-dominated asymmetric profile was observed for other high-efficiency devices (Supplementary Information S9).

Interestingly, the devices with intermediate diode efficiency, such as Device 3 ($\theta = 12°$) and Device 7 ($\theta = 19°$), also display a SQUID-like SIP, but with a slight modulation envelope (Fig. 3c and Supplementary Information S9). Analysis of their supercurrent distribution reveals a minor yet non-negligible bulk supercurrent component alongside the dominant asymmetric edge supercurrent (Fig. 3d). In stark contrast, junctions with negligible SDE efficiency, such as Device 4 and Device 9 ($\theta = 7°$ and $= 24°$), displayed a prominent decaying Fraunhofer-like envelope superimposed with small-period oscillations (Fig. 3e). The corresponding supercurrent distribution analysis confirmed that while some edge currents persist, the bulk contributions are now comparable to the edge supercurrent (Fig. 3f). The SIPs under applied field along different directions demonstrate the persistent edge supercurrent, displayed in

Supplementary Information S11.

Taken together, these results establish a clear correlation between the superconducting current pathway and the field-free SDE: an efficient SDE is observed only in junctions where the asymmetric supercurrent is strongly localized at the edges, while devices with a comparable bulk supercurrent exhibit a vanishing SDE. This pivotal finding suggests that the nonreciprocal transport in our junctions is fundamentally linked to these asymmetric edge supercurrents. The twist angle acts as a geometrical tuning knob for manipulating the spatial current distribution, transitioning the junction from an edge-dominated, highly nonreciprocal state to a bulk-dominated, reciprocal state.

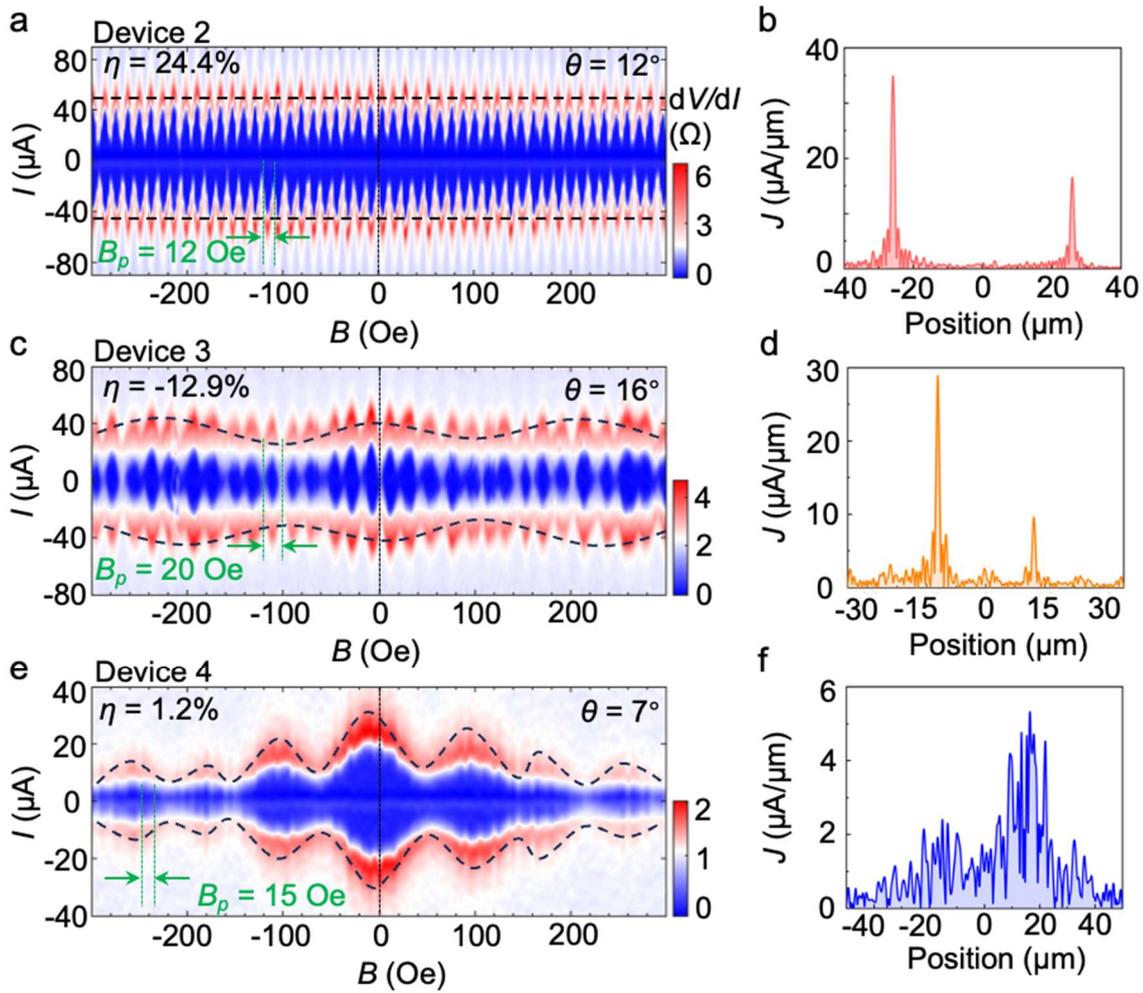

**Fig. 3 | Twist-angle dependence of superconducting interference patterns (SIPs) and the corresponding supercurrent distributions.** a, SIP of Device 2 ($\theta = 27°$), with period $B_p \approx 12$ Oe and diode efficiency $\eta = 24.4\%$. The interference envelope (dashed lines) shows no decay.

b, Supercurrent distribution of Device 2 extracted from (a), indicating predominantly edge-dominated transport. c, SIP of Device 3 ($\theta = 12°$), with a fast-oscillating component $B_p \approx 20$ Oe and $\eta = 12.9\%$. A slower envelope modulation is visible without clear decay. d, Corresponding supercurrent distribution of Device 3, showing finite bulk contribution. e, SIP of Device 4 ($\theta = 7°$), with $B_p \approx 15$ Oe and $\eta = 1.2\%$. The envelope exhibits noticeable decay. f, Corresponding supercurrent distribution of Device 4, indicating enhanced bulk contribution.

**Discussion and Summary**

Twist engineering provides direct control over the SDE in $FeTe_xSe_{1-x}$ / $NbSe_2$ junctions. The key mechanism is a twist-angle-dependent redistribution of supercurrents between bulk and edge channels, revealed by superconducting interference patterns. The strongest diode effect emerges when asymmetric edge currents dominate, while bulk transport suppresses nonreciprocity.

Bulk suppression likely arises from momentum mismatch between the $s\pm$-wave order of $FeTe_xSe_{1-x}$ and the $s$-wave order of $NbSe_2$, which cancels first-order Josephson tunneling and enables higher-order processes[31–33] (Figs. 4a and b, see Supplementary Information S12). Ferromagnetism in $FeTe_xSe_{1-x}$ (Supplementary Information S13) may provide an additional source of bulk suppression[41,42] (Supplementary Information S12), but how these ingredients combine with the systematic twist dependence remains unresolved. The microscopic origin of the edge currents—whether topological boundary modes[43,44] or ferromagnetic domain-wall states[45]—remains a central open question.

These findings establish twist-controlled heterostructures as a platform for large, field-free SDEs with record efficiency. By linking nonreciprocity to edge transport, pairing mismatch, and magnetism, they reveal a new pathway to high-performance superconducting diodes and provide a foundation for exploring how twist, topology, and magnetism conspire to produce emergent quantum phenomena.

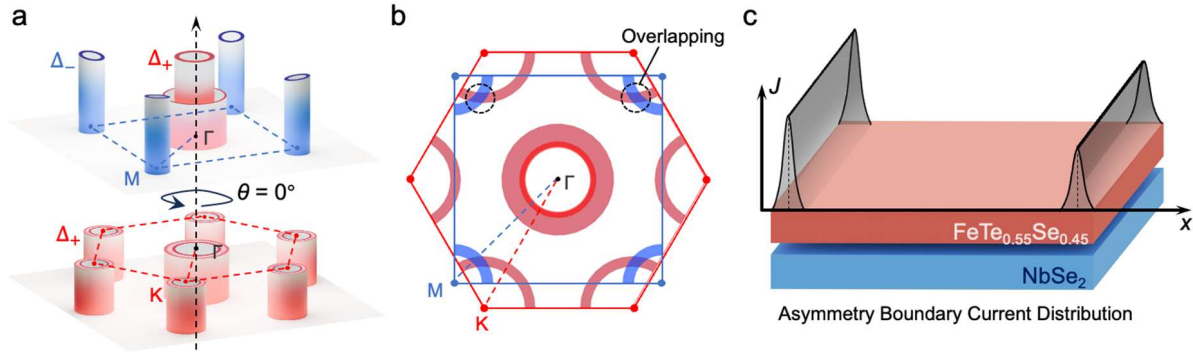

**Fig. 4 | Schematics of bulk supercurrent suppression and asymmetric edge transport.** a, Momentum mismatch between the $s\pm$ pairing of FeTe$_x$Se$_{1-x}$ and the $s$-wave order of NbSe$_2$ suppresses first-order Josephson tunneling through cancellation of opposite-sign order parameters. b, The schematic of the coupling between two different superconducting pairing gap with different signs. c, Asymmetric edge supercurrents in FeTe$_{0.55}$Se$_{0.45}$/NbSe$_2$ junctions, where boundary-localized states contribute strongly to the observed diode efficiency.

## METHODS

**Fabrication of devices**

Device fabrication began with patterning and depositing four Ti/Au (5 nm/50 nm) back electrodes by laser direct writing and magnetron sputtering. $FeTe_{0.55}Se_{0.45}$, $NbSe_2$, and h-BN nanoflakes were then prepared via mechanical exfoliation inside a glove box, where oxygen and water concentrations were maintained below 0.01 ppm to prevent oxidation. Using a dry-transfer technique, the $NbSe_2$ layer was placed to bridge two of the back electrodes, followed by stacking the $FeTe_{0.55}Se_{0.45}$ flake on top of $NbSe_2$ to contact the remaining electrodes. Finally, an h-BN capping layer was added to encapsulate the heterostructure and protect the device from degradation.

**Resistance measurements**

Temperature-dependent resistance and magnetoresistance of the Josephson junctions were measured using standard four-probe techniques in a physical property measurement system (PPMS, Quantum Design), with an a.c. excitation current of 5 μA.

**SDE measurements**

Prior to cooling from 300 K to 2 K, the magnetic field was ramped down from 30,000 Oe to zero in oscillation mode to suppress trapped flux in the superconducting magnet. V–I characteristics were measured at low temperatures using the PPMS Electronic Transport Option. The bias current was swept sequentially from zero to a positive maximum (0 → P), back to zero (P → 0), then to a negative maximum (0 → N), and finally back to zero (N → 0). Critical currents (Ic) were extracted from peaks in the differential resistance (d$V$/d$I$–$I$) curves (Figs. 2a–c). To quantify the SDE, $I_c^+$ and $I_c^-$ were defined as the critical currents obtained in the "0 → P" and "0 → N" sweeps, respectively, consistent with the principle of entropy increase. The temperature dependence of the SDE was determined by measuring $V$–$I$ curves between 2 K and 7 K under zero field.

**Measurements of interference patterns**

SIPs were obtained from $V$–$I$ curves measured in a standard four-terminal configuration at discrete $B$ with step sizes of 2 Oe or 5 Oe, depending on the oscillation period of the Josephson critical current. The field was applied in-plane, perpendicular to the $c$-axis of both FeTe$_{0.55}$Se$_{0.45}$ and NbSe$_2$. Sweeps from positive to negative $B$ were defined as negative sweeps, and vice versa for positive sweeps. The maximum applied current was always set below 100 $\mu$A, determined by the device $I_c$. We systematically investigated four types of SIP dependencies:
- B-range dependence: SIPs were recorded under different maximum field ranges.
- Field-direction dependence: SIPs were measured with varying in-plane field orientations.
- Temperature dependence: SIPs were collected between 2 K and 7 K, spanning the $T_c$ of NbSe$_2$.
- Field-cooling dependence: SIPs were obtained after warming to 18 K and cooling to 2 K

under finite applied $B$.

**Acknowledgements**
We sincerely thank G. Gu from Brookhaven National Laboratory for providing high-quality Fe(Te,Se) single crystals (prior to 2018); X. C. Xie, H. Jiang, Q. Yan and C. Chen from Fudan University, as well as K. T. Law, Z. T. Sun from Hong Kong University of Science and Technology, for their valuable theoretical discussions; Y. Huang and J. Ren from Shanghai Advanced Research Institute, CAS, for their assistance with the PPMS measurements; D. Pan, F. He, Y. Yin and W. Yang from Institute of Semiconductors, CAS, together with C. Cheng, from Shanghai Jiao Tong University, for their helpful assistance with the TEM characterization; F. Dai and S. Cui for their assistance with the EBSD characterization. H. D. acknowledges support from the New Cornerstone Science Foundation (No. 23H010801236), Innovation Program for Quantum Science and Technology (No. 2021ZD0302700). B. L. acknowledges support from the Ministry of Science and Technology of China (2023YFA1407400), the National Natural Science Foundation of China (12374063), the Shanghai Natural Science Fund for Original Exploration Program (23ZR1479900), and the Cultivation Project of Shanghai Research Center for Quantum Sciences (Grant No. LZPY2024). X. Z. acknowledges support from the National Natural Science Foundation of China (62304166). We acknowledge the technical support provided by the BL09U (31124.02.SSRF.BL09U) beamline at the Shanghai Synchrotron Radiation Facility (SSRF) for conducting the PPMS measurements.